# Inferência Baseada em Magnitudes na investigação em Ciências do Esporte. A necessidade de romper com os testes de hipótese nula e os valores de *p*

Rui MARCELINO[1,2,3]

Bruno Natale PASQUARELLI[4]

Jaime SAMPAIO[1,2]

1. Research Centre in Sports Sciences, Health Sciences and Human Development, CIDESD, CreativeLab Research Community, Portugal;
2. University of Trás-os-Montes and Alto Douro, UTAD, Portugal;
3. University Institute of Maia, ISMAI, Portugal;
4. Universidade Estadual de Campinas, Campinas-SP, Brasil

**Resumo**

As investigações em ciências do esporte sustentam-se frequentemente em inferências baseadas na declaração de um valor estatisticamente significativo, ou não significativo, com base no valor de *p* que deriva dos testes de hipótese nula. Considerando que os estudos são iminentemente amostrais, o recurso aos testes de hipótese nula apenas possibilita estimar os valores verdadeiros (população) das estatísticas utilizadas. Contudo, tem crescido a evidência, em diversas áreas do conhecimento, de que esta abordagem origina frequentemente interpretações confusas e até erradas (7). Para ultrapassar esta limitação têm surgido recentemente recomendações no sentido de sustentar as análises estatísticas com abordagens que recorram a interpretações mais intuitivas e mais práticas, baseadas sobretudo nas magnitudes (certezas/incertezas) dos valores verdadeiros encontrados. Com o intento de fornecer pistas alternativas aos desenhos metodológicos recorrentemente utilizados na investigação em ciências do esporte, neste trabalho procuraremos i) enunciar sucintamente algumas das fragilidades associadas aos testes de hipótese nula sustentados no valor de *p*; ii) refletir sobre as implicações da utilização da significância prática/clinica em oposição à significância estatística; iii) apresentar propostas de utilização das técnicas de inferências baseadas na magnitude, particularmente na visualização e interpretação dos resultados; iv) apresentar as principais limitações do uso das inferências baseadas em magnitudes. Assim, neste artigo de atualização desencoraja-se, de forma sustentada e fundamentada, o uso dos testes de significância baseados apenas no conceito de hipótese nula. Em alternativa, propõe-se a utilização de métodos de inferências baseados em magnitudes por possibilitarem interpretações dos efeitos práticos/clínicos dos resultados obtidos.

**PALAVRAS-CHAVE:** Estatísticas progressivas; análises quantitativas; desenho metodológico; significância clínica/prática; Inferência qualitativa

## Introdução

A investigação científica em Ciências do Esporte sustenta-se frequentemente nos conceitos de inferência estatística formulados por Neyman-Person[a], - que se caracterizam pela aceitação ou não aceitação da hipótese nula com base no valor de p – para produzir conhecimentos relevantes adstritos às matérias em estudo. Por convenção, esses novos conhecimentos são publicados sob a forma de artigos em revistas científicas, nos quais se apresentam também todas as evidências que os suportam, tentando convencer os leitores de que os resultados apresentados são verdadeiros. Contudo, em 2005, o estatístico John Iaonnidis (1) publicou um artigo com grande impacto na comunidade científica (citado mais de 1800 vezes e visto mais de 1,4 milhões de vezes), no qual argumenta que a maioria dos resultados científicos publicados são falsos, apontando como principais problemas: i) a investigação publicada ser uma seleção enviesada da totalidade de investigação produzida; ii) as análises de dados e os resultados apresentados serem também seletivos e enviesados; iii) em muitas áreas de investigação, os estudos raramente serem replicados, pelo que persistem as falsas conclusões. De acordo com este autor, as conclusões apresentadas nas investigações científicas são fundamentadas quase exclusivamente nos resultados estatisticamente significativos (obtidos através de decisões baseadas no valor de $p$).

---

[a] Apesar dos conceitos de 'Hipótese Nula' (representado por $H_0$), de 'significância', de 'graus de liberdade', assim como de distinção entre população e amostra serem da autoria do Sir Ronald Fisher (1890-1962), tal como a sugestão do valor arbitrário de $p < 0.05$ para tomar decisões sobre a $H_0$ (R. Fisher, 1925), foram os matemáticos Jerzy Neyman e Egon Pearson que sugeriram pela primeira vez os conceitos de inferência estatística, assim como demonstraram os procedimentos específicos e a lógica de interpretação dos testes de hipótese nula, que perduram até aos dias de hoje (Neyman, 1933). Foram igualmente estes últimos autores que introduziram a 'Hipótese Alternativa' (representada por $H_1$), 'statistical power', falsos positivos e falsos negativos.

Entretanto, nos últimos anos os testes de significância baseados na 'hipótese nula' (TSHN) têm sido questionados ao ponto de algumas revistas científicas não aceitarem a submissão de artigos com este tipo de tratamento estatístico (2) ou então recomendarem informações adicionais utilizando análises alternativas (3-6).

Tendo em vista que o conhecimento atualiza-se em todas as áreas do saber a medida que cresce o número de trabalhos publicados, é de se esperar que os métodos estatísticos, que são também objeto de constante investigação, caminhem igualmente neste sentido. Dessa forma, alguns posicionamentos em áreas que utilizam maioritariamente variáveis numéricas, tais como Medicina, Enfermagem, Psicologia ou Fisioterapia, tenham sido feitos com respeito a atualização no método estatístico a ser utilizado (7-12). Nestes documentos os autores sugerem mudanças nos paradigmas usados para a análise dos dados, indicando estratégias alternativas às convencionais, ou seja, alternativas aos TSHN (7-12). Existindo na área de Ciências do Esporte um enorme número de variáveis quantitativas, logo numéricas, urge igualmente procurar seguir aquelas que são correntemente enunciadas como as "boas práticas" na análise estatística de dados quantitativos.

Desde maio de 1997 que Will Hopkins mantém o seu site pessoal atualizado com aspetos relevantes para o tratamento estatístico de dados esportivos, numa coletânea intitulada *A New View of Statistics*. Foi a partir do ano de 2000 que surgiram pela primeira vez neste espaço abordagens alternativas e complementares aos TSHN (13). Foi com base na informação aqui sintetizada que foi redigido por Will Hopkins e colaboradores (4) **um** dos primeiros artigos científicos dedicado a analisar esta problemática nas Ciências do Esporte , tendo sido publicado em 2009 na revista *Medicine & Science in Sports & Exercise*. Os autores apresentaram uma proposta de avanço em análise estatística na área de Ciências do Esporte com o intuito de promover o debate e construir um novo caminho para a análise de dados. O que os autores chamaram de *Estatística Progressiva* foi o modelo de Inferência Baseada em Magnitude (MBI, do termo em inglês *Magnitude-Based Inference*). Com esta abordagem pretende-se determinar a diferença entre grupos com base na inferência dos efeitos, ou seja, na diferença prática ou clínica, e não na significância estatística baseada na hipótese nula.

Entretanto, a sugestão da aplicação da IBM nas Ciências do Esporte gerou um grande debate da comunidade científica. Welsh e Knight (14) lançaram mão de críticas a este modelo, alegando não haver suporte matemático suficiente para ser validado. Nesta critica tentou demonstrar-se que i) os cálculos da MBI não derivam diretamente dos intervalos de confiança mas sim dos valores de $p$ para alguns testes particulares; e que ii) a MBI é menos conservadora do que as inferências tradicionais uma vez que altera a hipótese nula e utiliza valores de $p$ considerados para um lado da distribuição (uma cauda), no lugar de considerar os dois lados da distribuição (duas caudas) (14).

Outros pesquisadores foram envolvidos na discussão, que não se findou até o momento. Entretanto, como sempre ocorre quando um novo paradigma viola as evidências do paradigma vigente, chamado de Ciência Normal (15), a discussão prossegue. É uma discussão muito interessante de acompanhar e que coloca em oposição cientistas de áreas de conhecimento muito diversas: matemáticos teóricos, especialistas em estatística aplicada, assim como pesquisadores com grande experiência no contato com a área de atuação das suas áreas profissionais, têm vindo progressivamente a enumerar argumentos que contribuem para uma análise crítica desta questão. Atualmente, pelo menos assim parece, os argumentos a favor da aplicação das técnicas associadas à MBI, para o estudo de fenômenos humanos, têm-se superiorizado de forma muito vincada, apresentando de forma inequívoca as fragilidades dos TSHN.

Assim, neste trabalho procuraremos i) enunciar sucintamente algumas das fragilidades associadas aos testes de hipótese nula sustentados no valor de *p;* ii) refletir sobre as implicações da utilização da significância prática/clinica em oposição à significância estatística; iii) apresentar propostas de utilização das técnicas de MBI, particularmente na visualização e interpretação dos resultados: iv) apresentar as limitações da MBI. Tudo isto tendo sempre em linha de pensamento, obviamente, os problemas específicos das Ciências do Esporte.

**Testes de hipótese nula e valores de *p***

Apesar da controvérsia acerca da possibilidade dos testes de hipótese nula responderem às questões da investigação ter-se iniciado nos anos 60 (16), a verdade é que este tem sido o paradigma dominante na investigação em Esporte. Colquhoun (11) aponta que abordagens que usam o *valor de p* com nível de significância de 5% podem estar no mínimo 30% das vezes enganadas sobre o real valor das diferenças. Recentemente a revista *Nature* publicou um posicionamento no qual assume que os TSHN não são tão fiáveis quanto pensam os pesquisadores (7). Na fala do físico e estatístico Steven Goodman, da Escola de Medicina da Stanford University, *"o valor de p não foi concebido para ser usado da maneira como é usado hoje"*. Acrescenta ainda que uma mudança de filosofia em análise estatística de dados é fundamental, pois evitará que cientistas percam informações importantes ou se guiem por falsos-positivos (7).

Cohen [7], Kline (17) e Cummings (18) apresentam explicações detalhadas sobre os problemas/limitações dos testes de hipótese nula. Sinteticamente,

> -valor de *p* não é a probabilidade da hipótese nula;
> -rejeitar a hipótese nula não prova que a hipótese alternativa é verdadeira;
> -não rejeitar (aceitar) a hipótese nula não prova que a hipótese alternativa é falsa;
> -o valor de *p* não dá indicação do tamanho do efeito.

De acordo com Gardner e Altman, utilizar o valor arbitrário de 5% como nível de significância estatística para definir duas possibilidades – significativo ou não significativo – "não é útil e encoraja o pensamento preguiçoso" (19). Independentemente da interpretação do valor de *p,* os testes de hipóteses são ilógicos, porque a hipótese nula de não relação ou não diferença é sempre falsa, uma vez que não há verdadeiros efeitos zero na natureza (3).

Na formulação de um problema ou de uma questão de pesquisa, os investigadores têm normalmente fundamentos para acreditar que os efeitos serão diferentes de zero. Assim, mais relevante do que saber se o efeito existe é perceber quão grande ele é.

Infelizmente, o valor de *p* por si só não fornece qualquer tipo de informação sobre a direção ou tamanho do efeito (20). Dependendo, por exemplo, do tamanho e da variabilidade da amostra, um resultado estatístico com $p<0,05$ pode representar um efeito que é clinicamente irrelevante. Ao contrário, um resultado não significativo ($p>0,05$) não implica necessariamente que o efeito não deva ser tido em consideração, isto porque a combinação de amostras de tamanhos reduzidos com elevadas variabilidades pode mascarar efeitos importantes (21). Para ilustrar esta afirmação, a Figura 1, adaptada de Cumming (22) apresenta exemplos de como o TSHN analisam os dados quando o experimento é repetidos 25 vezes. Para demonstrar até que ponto o valor de *p* pode falhar na identificação de diferenças entre grupos, este autor criou dados simulados de dois grupos independentes com as seguintes características: cada um com um total de 32 elementos; os valores testados seguem a distribuição normal nos dois grupos; um dos grupos apresenta uma média superior em 10 unidades de medida. Ou seja, a diferença de médias na população é de 10. Estabelecidos estes parâmetros, o autor testou as diferenças de médias em amostras aleatórias de 20 elementos em cada um dos grupos. Assim, como se pode comprovar na Figura 1, apesar de a diferença da população (conhecida à partida) ser efetiva entre os dois grupos, quando calculadas as diferenças amostrais e analisados os valores de *p,* dos 25 experimentos, apenas em 11 testes foram encontrados valores de $p<0.05$, logo indicadores de diferença entre os grupos (Experimentos nº: 1, 3, 6, 8, 11, 12, 14, 19, 20, 21 e 24). Nos restantes casos, seriamos levados a concluir que não há diferenças estatisticamente significativas entre os dois grupos; apesar de as diferenças existirem. Este exercício muito simples, permite concluir que na realidade a capacidade do valor de *p* detetar diferenças entre grupos deve-se, em grande medida, a sorte na seleção dos elementos constituintes das amostras. Sabendo nós que muito dificilmente nas investigações na Ciências do Esporte teremos acesso à totalidade da população, tendo que investigar necessariamente com valores provenientes de amostras, torna-se evidente a fragilidades em fazer inferências apenas com os valores de *p*. Na mesma figura apresenta-se igualmente a média das diferenças juntamente com o Intervalo de Confiança a 95%, que poderá figurar-se como uma alternativa à interpretação exclusiva dos valores de *p*. Mais a frente neste artigo apresentaremos possíveis leituras destes valores.

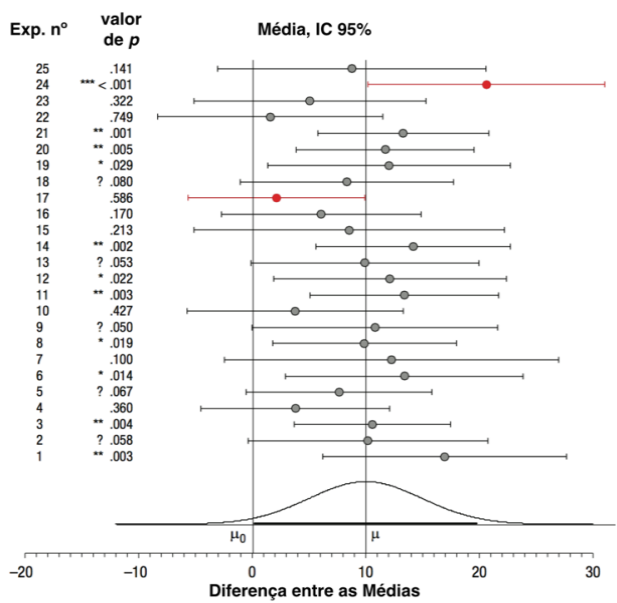

**Figura 1.** Resultados simulados de 25 repetições de um experimento (número à esquerda do gráfico). Cada experimento resulta de duas amostras independentes provenientes de uma população com $n=32$; as amostras estão normalmente distribuídas com $\alpha = 20$ e médias populacionais diferem em $\mu = 10$. Para cada experimento, as diferenças entre as médias (círculo) e o IC 95% destas diferenças estão representados pelas barras. A curva é a distribuição dos dados respectivos das diferenças; a linha em negrito corresponde abaixo do gráfico corresponde ao intervalo de 95%. Os valores de p correspondem a: *$0.01 < p < 0.5$, **$0.001 < p < 0.01$, ***$p < 0.001$; o ponto de interrogação indica que $0.05 < p < 0.10$. Figura adaptada de Cumming (22).

**Significância estatística vs. Significância prática/clínica com recurso à inferência estatística**

Na investigação em Esporte, tal como em todas as outras áreas de conhecimento, deve haver uma distinção clara entre a significância estatística e a prática/clínica (23). A *significância estatística* – principal foco dos testes de hipótese nula – está relacionada com a possibilidade dos resultados serem atribuídos à variabilidade da amostra, enquanto que a *significância prática/clinica* está relacionada como a utilidade dos resultados.

Considerando, a título de exemplo, o teste acerca da eficácia de um programa de desenvolvimento de força, os investigadores poderão questionar-se: "o programa funciona?" e "quão bem funciona?", ou seja, "quais os ganhos esperados ao nível do aumento da força?". Para além destas questões deverão igualmente questionar-se: "qual é a precisão da estimativa sobre os ganhos ao nível da força?".

Um reduzido tamanho da amostra, ou uma grande variabilidade individual, poderão resultar numa incerteza acerca da eficácia do programa de treino, apesar de, hipoteticamente, encontrarmos uma grande diferença nas médias das performances entre os grupos intervencionados e não intervencionados. Os intervalos de confiança apresentam-se como uma medida capaz de captar a imprecisão da medida (ou seja, do tamanho do efeito), representando uma amplitude plausível na qual o valor verdadeiro (mas desconhecido) da população se situa (24).

A questão mais difícil de responder será, eventualmente: "a intervenção – aplicação do programa de desenvolvimento de força – é rentável, i.e. custo-efetiva?"

Apesar da dificuldade em responder de forma inequívoca a esta questão, ela deverá ser considerada na investigação. Uma possibilidade de avaliar a eficácia de uma intervenção poderá ser através da definição prévia da diferença clínica/prática mínima para considerar a intervenção efetiva. Esta diferença é frequentemente considerada como sendo equivalente a um tamanho do efeito de 0.20 (25).

Em 2001, Shakespeare et al. (20), num artigo publicado na revista *The Lancet*, defenderam que o processo de interpretação dos resultados da investigação médica e a sua tradução nas práticas clínicas deveria ser efetuado com recurso a intervalos de confiança, curvas de significância clínica e análises de risco-benefício, e não apenas nos valores de *p*. Estas mesmas estratégias, particularmente a utilização de intervalos de confiança para os tamanhos dos efeitos (como apresentados na Figura 2), embora não garanta a perfeita interpretação dos resultados, têm sido recorrentemente referidas como potenciadora do fornecimento de informação acerca da significância prática/clínica das investigações (3-6, 26).

Na utilização das MBI é frequente o recurso a transformações logarítmicas para lidar com a não-uniformidade, quer dos efeitos quer dos erros, frequente nos dados recolhidos em investigações esportivas. Esta não-uniformidade nos dados brutos (do inglês *raw data*) em modelos lineares pode levar à produção de estimativas e limites de confiança incorrectos (4), pelo que sempre que se verificar que existe heterocedasticidade regular, quando a variância é proporcional à medida do tratamento, deve recorrer-se à transformação logarítmica, sendo esta geralmente eficiente (27).

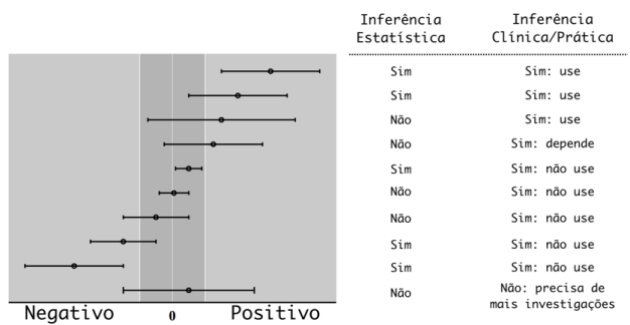

**Figura 2**. Exemplos de resultados baseados na diferença entre as médias e IC 90% com inferência ao efeito clínico/prático e estatístico de significância baseada na hipótese nula. Adaptado de Batterham e Hopkins (26)

### Inferência Baseada em Magnitude em Ciências do Esporte: Propostas

A MBI é um modelo mais intuitivo e possibilita ao pesquisador uma análise muito mais interpretativa e menos dicotômica, quando comparado aos TSHN. As inferências qualitativas deste modelo possibilitam saber em que direção se verificam as diferenças: positivo/benéfico, sem efeito/irrelevante, negativo/maléfico; e em qual magnitude tem estas variáveis de acordo com a variabilidade encontrada (ver Figura 3).

Adicionalmente a informação da magnitude do efeito, pode-se colocar o efeito prático/clínico destas diferenças. Conforme citado anteriormente, a ideia deste conceito é a de fornecer uma análise qualitativa da real diferença entre as variáveis de diferentes grupos ou do real efeito de um tratamento ou programa de treinamento. Na Figura 4, apresentam-se os mesmos resultados da Figura 2, mas desta vez incluindo informação acerca da probabilidade/chance de novas análises aleatórias dos mesmos dados, originarem valores negativos, triviais ou positivos nos tamanhos do efeito. Apresentam-se também propostas para a interpretação/inferência qualitativa prática/clinica do estudo em questão. Assim, se observarmos o valor apresentado no topo superior do gráfico, e ainda a respeito do hipotético teste acerca da eficácia de um programa de desenvolvimento da força, somos levados a concluir que a eficácia neste grupo é muito elevada, sendo o efeito "quase certamente Positivo". Podemos igualmente saber que se recolhermos aleatoriamente dados de um dos elementos constituintes do grupo, a chance de encontrarmos valores inferiores após a aplicação do programa de treinamento é de 0,01%; a chance de não encontrarmos diferenças (ou seja, diferenças triviais) é de 1% e a chance de encontrarmos valores superiores após o programa de treinamento é de 99%. Continuando a leitura do gráfico, apesar de os dois valores seguintes (lendo de cima para baixo) serem Positivos, observamos que um é "provavelmente" e o outro "possivelmente". No caso do "Possivelmente Positivo", para a maioria dos elementos constituintes deste grupo (65%) o programa de treinamento foi benéfico, ou seja, permitiu aumentos dos níveis de força. Contudo, sabemos igualmente que a probabilidade de encontrar elementos do grupo que diminuíram os seus valores de força após a aplicação do programa é de 2% e a de encontrar elementos sem diferenças após a aplicação é de 33%.

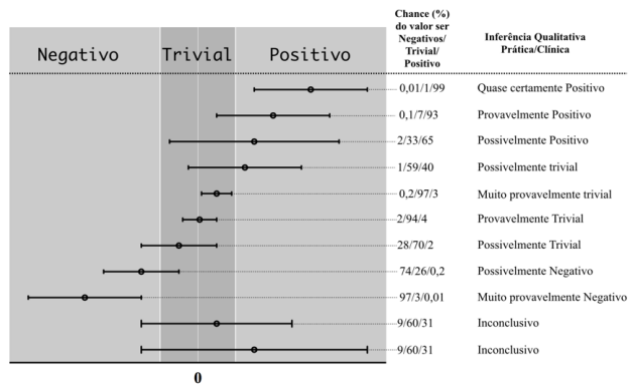

**Figura 3**. Tamanho/Magnitude do efeito de Cohen em unidades estandardizadas. As diferenças entre as médias estão representadas pelo círculo e o IC 95% destas diferenças estão representados pelas barras. No exemplo da figura a direção da média em relação ao ponto "0" indica se houve ou não efeito do treinamento. : <0,2 , "efeito trivial ou nulo"; 0,2 a 0,6 , "efeito pequeno"; 0,6 a 1,2 , "efeito moderado"; 1,2 a 2,0 , "efeito grande"; >2,0 , "efeito muito grande".

*Tamanho do Efeito*

Em estatística, o tamanho do efeito é uma medida quantitativa sobre "a força" de um fenómeno. O tamanho do efeito de Cohen é um dos principais indicadores de diferenças entre variáveis dependentes (efeito de um determinado treinamento no momento controle e após intervenção, por exemplo) ou independentes (diferença entre médias de dois grupos, por exemplo). Vários programas estatísticos permitem calcular os tamanhos do efeito associados aos diferentes tipos de comparações efectuadas, estando igualmente disponíveis on-line diversas planilhas (28) para os mesmos cálculos. Na Figura **4** o eixo "x" indica o tamanho ou magnitude do efeito (neste caso a média das diferenças em unidades estandardizadas), representado pelo símbolo posicionado no centro das barras; sendo que estas representativas do intervalo de confiança (IC). O IC de 95% (29) e 90% (4) são os mais usuais. A decisão entre utilizar um ou outro cabe aos pesquisadores e deve ser justificada de acordo com o tipo de variáveis que está tratando. A escolha do IC pode interferir nas inferências práticas/clínicas baseadas em magnitudes. Portanto, em estudos de revisão sistemática ou comparações entre amostras independentes, a utilização do IC95% pode ser mais honesta. Nos casos de inferências práticas de um mesmo grupo, como nas análises de desempenho no esporte, o IC90% parece ser mais aceito e utilizado. Os valores de corte para a interpretação qualitativa da magnitude do efeito (30) estão igualmente apresentados na Figura 4.

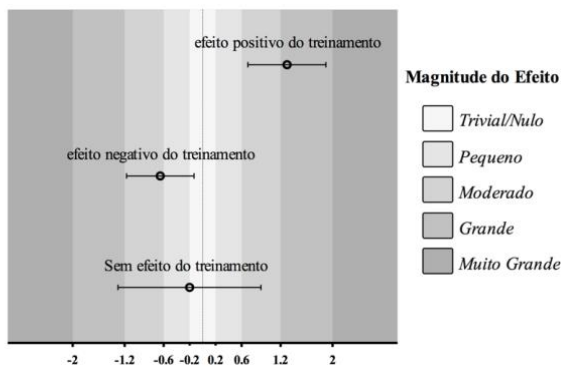

**Figura 4.** Tamanho/Magnitude do efeito de Cohen em unidades estandardizadas. Valores interpretativas para a inferência quantitativa prática/clinica dos tamanhos do efeito em função dos intervalos de confiança. Adaptado de Batterham e Hopkins (26).

Os valores apresentados nas Figuras poderão ser complementados com os elementos apresentados, a título de exemplo, na Tabela 1.

**Tabela 1.** Sugestão de formato de tabela para estudos em Ciências do Esporte.

|  | **Grupo 1 Ou Res. Pré (Média± DP)** | **Grupo 2 Ou Res. Pós (Média± DP)** | **Dif. Entre as médias; ± IC** | **%Dif.; ±%IC** | **Tamanho do Efeito** (Média (IC95 %) |
|---|---|---|---|---|---|
| Variável 1 | 2,9±0,5 | 3,9±1,2 | 1,0; ±0,8 | 2,1 %; ±3,6 % | 1,2 (0,6 a 2) |

É frequente que em estudos na área do esporte as amostras sejam pequenas. Tendo em vista esta questão, Weissegerber et al. (31) propuseram, em um artigo publicado na revista *PLoS Biology,* uma forma de representação gráfica dos resultados que permitam uma melhor visualização, permitindo análises individuais concomitantemente às análises dos grupos em estudo. Adicionalmente a proposta dos autores, cordialmente dispuseram na página da revista uma planilha em *Excel (Microsoft Office®)* na qual os gráficos podem ser gerados (31).

Geoff Cumming, um dos maiores críticos da utilização do TSHN, também sugeriu uma mudança de paradigma na representação dos dados, muito bem explicada no seu livro *Understanding the new statistics: Effect sizes, confidence intervals, and meta-analysis* (18), e também disponibilizou uma planilha em *Excel (Microsoft Office®)* para que sejam elaborados os gráficos (22, 29).

*Limitações do uso das MBI*

Apesar dos benefícios apresentados enquanto alternativa ao uso dos TSHN para estudar fenómenos esportivos, é necessário considerar algumas limitações do uso das MBI. De acordo com Schaik e Weston (32) é possível organizar estas limitações em aparentes e em substantivas. As limitações aparentes referem-se a aspectos que frequentemente são classificados como limitadores, mas que na verdade se refletem como vantajosos em oposição aos TSHN. Assim, os autores apresentam como limitação aparente o facto de os pesquisadores terem que tomar opções, ou aceitarem as escolhas recomendadas, relativamente a alguns parâmetros nas análises. Em cada análise é necessário estabelecer o tamanho do efeito mínimo para ser considerado importante (*smallest worthwhile change* ou *smallest important effect size*), as taxas de erro tipo I e de erro tipo II, assim como categorizar os valores das probabilidades quantitativas em descritores qualitativos (ver Figura 4). As opções a tomar (especialmente a definição do tamanho do efeito mínimo para ser considerado importante) no momento de efetuar as análises terão que decorrer da compreensão detalhada do fenômeno esportivo em estudo. Quando estas escolhas não coincidem na totalidade com os valores recomendados (30) recomenda-se uma argumentação fundamentada de forma credível, para convencer os revisores/editores/leitores (32). Contudo, é fácil compreender que são os pesquisadores quem melhor conhece os fenómenos em estudo, pelo que as opções serão muito fáceis de tomar e de fundamentar. No que diz respeito às limitações substantivas, são recorrentemente referenciados os argumentos defendidos por Welsh e Knight (14), segundo os quais a IBM inflaciona a probabilidade de detetar um efeito quando ele não existe. Todos estes argumentos têm sido contrariados com fundamentação estritamente matemática, através da discussão académica anteriormente referida.

O fato de a generalidade das inferências baseadas nas magnitudes serem produzidas com recurso a planilhas em Excel, sem recurso a software especificamente dedicado a estas análises, exige que os pesquisadores redobrem os cuidados na parametrização de todos os indicadores para que não se verifiquem erros de codificação ou a presença de fórmulas corrompidas ao longo das planilhas.

**Considerações Finais**

O presente estudo abre uma discussão sobre a utilidade destas novas ferramentas em análise estatística, visto que a maior clareza das informações pode diminuir o tempo para elucidação de um fenômeno e consequentemente pode-se avançar a novos problemas nas Ciências do Esporte. Nesta perspectiva, deve-se cada vez mais desencorajar o uso dos testes de significância baseados na hipótese nula. A utilização do método de inferências baseadas em magnitude sugere a utilização do efeito prático/clínico para comparação de dados numéricos, permitindo uma análise qualitativa dos resultados encontrados. Adicionalmente, a representação dos resultados deve seguir novas tendências para facilitar a compreensão e promover um maior entendimento das descobertas.

## Magnitude-Based Inferences in Sports Sciences Research. The need to break away from the null-hypothesis and the p values

**Abstract**

Research in Sports Sciences is supported often by inferences based on the declaration of the value of the statistic statistically significant or nonsignificant on the bases of a P value derived from a null-hypothesis test. Taking into account that studies are manly conducted in sample, the use of null hypothesis testing only allows estimating the true values (population) of the statistics used. However, evidence has grown in many areas of knowledge that this approach often leads to confusion and misinterpretation. To overcome this limitation they have recently emerged recommendations to support the statistical analysis with approaches that make use of more intuitive interpretations and more practical, especially based on the magnitudes (certainty / uncertainty) of the true values found. With the intent to provide alternative solutions to methodological designs recurrently used in research in sports sciences, this paper will seek to i) briefly spell out some of the weaknesses associated with the null hypothesis tests based in the P value; ii) reflect on the implications of the use of practical/clinical significance as opposed to statistical significance; iii) submit proposals for use the inferences based on the magnitude, particularly in the visualization and interpretation of results; iv) present and discuss the limitations of magnitude-based inference. Thus, this update article discourages, in a sustained-based, the use of significance tests based only on the concept of null hypothesis. Alternatively, it is proposed to use methods of inference based on magnitudes as they allow interpretations of the practical/clinical effects results obtained.

**Keywords:** Progressive statistics; quantitative analysis; clinical significance; evidence based practice; qualitative inference

**Referências**

1. Ioannidis J. Why most published research findings are false. PLoS Med. 2005;2(8):e124.
2. Trafimow D. Editorial. BASP. 2014;36(1):1-2.
3. Batterham A, Hopkins W. The case for magnitude-based inference. Med Sci Sports Exerc. 2015;47(4):885.
4. Hopkins W, Marshall S, Batterham A, Hanin J. Progressive statistics for studies in sports medicine and exercise science. Med Sci Sports Exerc. 2009;41(1):3-13.
5. Pyne D. Improving the Practice of Sports Science Research. Int J Sports Physiol Perform. 2014;9(6):899.
6. Winter E, Abt G, Nevill A. Metrics of meaningfulness as opposed to sleights of significance. J Sports Sci. 2014;32(10):901-2.
7. Nuzzo R. Scientific method: statistical errors. Nature. 2014;506(7487):150-2.
8. Sedgwick P. Statistical Question Clinical Significance Versus Statistical Significance. Brit Med J. 2014;348.
9. Savitz D. Is Statistical Significance Testing Useful in Interpreting Data. Reprod Toxicol. 1993;7(2):95-100.
10. Eken C. Statistical or clinical significance? A critical point in interpreting medical data. Am J Emerg Med. 2007;25(5):589-.
11. Colquhoun D. An investigation of the false discovery rate and the misinterpretation of p-values. R Soc open sci. 2014;1(3):140216.
12. Page P. Beyond statistical significance: clinical interpretation of rehabilitation research literature. International journal of sports physical therapy. 2014;9(5):726-36.
13. Hopkins W. A New View of Statistics: Internet Society for Sport Science; 2000 [Available from: http://www.sportsci.org/resource/stats/.
14. Welsh A, Knight E. "Magnitude-based inference": a statistical review. Med Sci Sports Exerc. 2015;47(4):874-84.
15. Kuhn T. A estrutura das revoluções cientificas: Perspectiva; 2001.
16. Meehl P. Theory-Testing in Psychology and Physics: A Methodological Paradox. Philos Sci. 1967;34(2):103.
17. Kline R. Beyond Significance Testing: Reforming Data Analysis Methods in Behavioral Research. Washington, DC: American Psychological Association; 2004.
18. Cumming G. Understanding the new statistics: Effect sizes, confidence intervals, and meta-analysis. New York, NY US: Routledge/Taylor & Francis Group; 2012.
19. Gardner M, Altman D. Confidence-Intervals Rather Than P-Values - Estimation Rather Than Hypothesis-Testing. Brit Med J. 1986;292(6522):746-50.
20. Shakespeare T, Gebski V, Veness M, Simes J. Improving interpretation of clinical studies by use of confidence levels, clinical significance curves, and risk-benefit contours. Lancet. 2001;357(9265):1349-53.
21. Cohen J. The Earth Is Round (P-Less-Than.05). Am Psychol. 1994;49(12):997-1003.
22. Cumming G. The New Statistics: Why and How. Psychol Sci. 2014;25(1):7-29.
23. Ferrill M, Brown D, Kyle J. Clinical versus statistical significance: interpreting P values and confidence intervals related to measures of association to guide decision making. Journal of pharmacy practice. 2010;23(4):344-51.
24. Coulson M, Healey M, Fidler F, Cumming G. Confidence intervals permit, but do not guarantee, better inference than statistical significance testing. Front Psychol. 2010;1(26).
25. Cohen J. A power primer. Psychol Bull. 1992;112(1):155-9.
26. Batterham A, Hopkins W. Making Meaningful Inferences About Magnitudes. Int J Sports Physiol Perform. 2006;1(1):50-7.
27. Sedgwick P. Log transformation of data. Brit Med J. 2012;345:2.
28. Hopkins W. A Spreadsheet to Compare Means of Two Groups 2007 [22-3]. Available from: http://sportsci.org/2007/inbrief.htm - xcl2.
29. Cumming G. The New Statistics: Estimation for better research 2012 [Available from: http://www.latrobe.edu.au/psy/research/cognitive-and-developmental-psychology/esci.
30. Hopkins W. A Scale of Magnitudes for Effect Statistics SportScience2002 [Available from: http://www.sportsci.org/resource/stats/effectmag.html.

31. Weissgerber T, Milic N, Winham S, Garovic V. Beyond bar and line graphs: time for a new data presentation paradigm. PLoS Biol. 2015;13(4):e1002128.
32. Schaik P, Weston M. Magnitude-based inference and its application in user research. Int J Hum Comput. 2016;88:38-50.